\documentclass[conference]{IEEEtran}
\IEEEoverridecommandlockouts
\usepackage[belowskip=-15pt,aboveskip=0pt]{caption}

\usepackage{bm}
\usepackage{algorithm}
\usepackage{algorithmic}
\usepackage[utf8]{inputenc}
\usepackage{multirow}
\usepackage{colortbl,booktabs}
\usepackage{environ}
\usepackage{float}
\usepackage{color,soul}
\usepackage{enumitem}
\usepackage{amsmath}
\usepackage{graphicx}
\graphicspath{{figs/}}
\usepackage{subfig}
\usepackage{float}
\usepackage{amsmath, amsthm, amsfonts, mathtools}
\usepackage{bm}
\usepackage{textcomp}
\usepackage{xcolor}
\usepackage{float}
\usepackage{color}
\usepackage{diagbox}
\usepackage{amsmath}
\usepackage{array}
\usepackage{makecell}
\usepackage{stfloats}
\usepackage{placeins}

\makeatletter
\newcommand{\thickhline}{%
    \noalign {\ifnum 0=`}\fi \hrule height 1pt
    \futurelet \reserved@a \@xhline
}

\newcolumntype{"}{@{\hskip\tabcolsep\vrule width 1pt\hskip\tabcolsep}}
\makeatother

\newcolumntype{[}{@{\vrule width 1pt\hspace{6pt}}} \newcolumntype{]}{@{\hspace{6pt}\vrule width 1pt}} \newcolumntype{!}{@{\hskip\tabcolsep\vrule width 1pt\hskip\tabcolsep}}

\begin{document}

\title{SoftmAP: Software-Hardware Co-design for Integer-Only Softmax on Associative Processors}

\author{
    Mariam Rakka, Jinhao Li, Guohao Dai, Ahmed Eltawil, Mohammed E. Fouda, and Fadi Kurdahi%
    \thanks{M. Rakka, and F. Kurdahi are with the University of California, Irvine, CA, USA (Email: mrakka@uci.edu).}%
    \thanks{J. Li and G. Dai are with Shanghai Jiao Tong University, Shanghai, China.}%
    \thanks{A. Eltawil is with King Abdullah University of Science and Technology (KAUST), Thuwal, Saudi Arabia.}%
    \thanks{M. Fouda is with Rain Neuromorphics Inc., San Francisco, CA, USA.}%
    \thanks{This work has been partially supported by King Abdullah University of Science and Technology CRG program under grant number:  URF/1/4704-01-01.}
}

\maketitle

\begin{abstract}
  Recent research efforts focus on reducing the computational and memory overheads of Large Language Models (LLMs) to make them feasible on resource-constrained devices. Despite advancements in compression techniques, non-linear operators like Softmax and Layernorm remain bottlenecks due to their sensitivity to quantization. We propose SoftmAP, a software-hardware co-design methodology that implements an integer-only low-precision Softmax using In-Memory Compute (IMC) hardware. Our method achieves up to three orders of magnitude improvement in the energy-delay product compared to A100 and RTX3090 GPUs, making LLMs more deployable without compromising performance.
\end{abstract}


\maketitle



\section{Introduction}
Large Language Models (LLMs) are gaining traction for demonstrating great performance in solving various complex real-world tasks, and paving the way towards Artificial General Intelligence (AGI) \cite{zhao2023survey, bubeck2023sparks}. The last 30 years have witnessed a tremendous increase in the inference compute capabilities (up to 7 orders of magnitude) of LLMs \cite{epoch2023pcdtrends}. LLMs follow the scaling laws for transformer-based models: their performance can predictably increase with a higher volume of data and a larger model size. This scalability, however, renders LLMs resource-hungry, hence confining their deployment, especially on resource-limited edge devices \cite{xu2024survey, kaplan2020scaling, kundu2024efficiently}.

Compression is used to alleviate the computational/memory overheads of LLMs~\cite{li2024large}. Quantization amongst other compression techniques (like pruning/knowledge distillation) is a popular topic. Quantization is mapping values in a large set to their quantized counterparts in a smaller set. Quantizing weights and activations in LLMs reduces memory footprint and speeds up the computation. For example, quantizing weights and activations from FP16 to INT8 decreases GPU memory usage by half and increases the throughput of performing matrix multiplications by up to $2\times$ \cite{wang2024model, xiao2023smoothquant}. LLM quantization frameworks like SmoothQuant, QLoRa, ZeroQuant, and AWQ \cite{xiao2023smoothquant,dettmers2024qlora, yao2022zeroquant, lin2023awq, li2024fastefficient2bitllm} have achieved reductions in computational and memory overheads of LLMs by quantizing weights. 
Beyond weights, quantizing activations in LLMs remains a challenge given that Softmax, a chief part of the attention mechanism, is sensitive to quantization \cite{pandey2023softmax, stevens2021softermax}. 

\begin{figure}[t!]
\centering
\includegraphics[width=0.9\columnwidth]{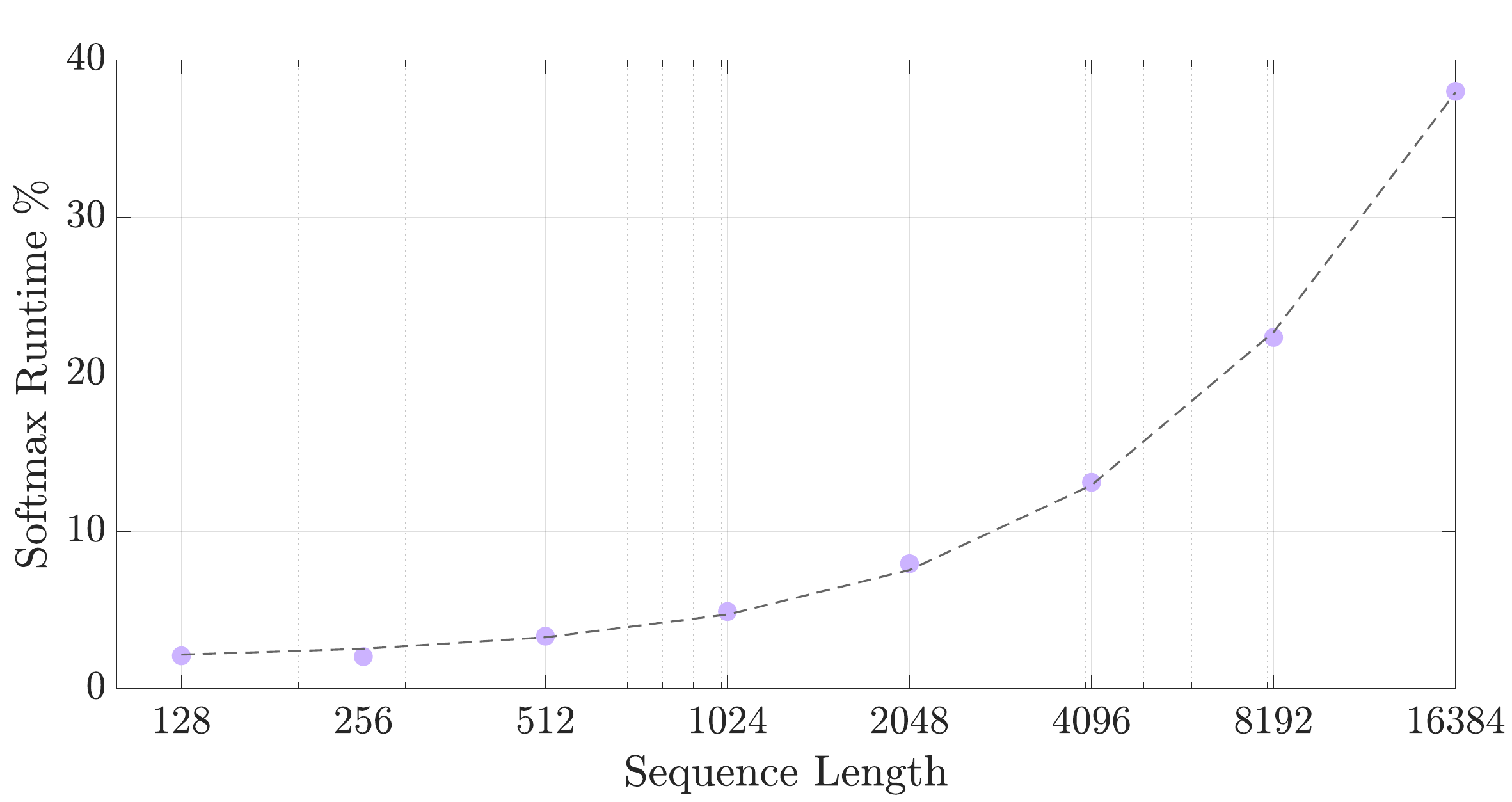}
\caption{Softmax runtime proportion for Llama2-7b on A100 (80GB). Softmax contributes up to 38\% of the run time for longer sequence lengths.}
\label{runtime_breakdown}
\end{figure}

\begin{figure}[t!]
\centering
\includegraphics[width=1\columnwidth]{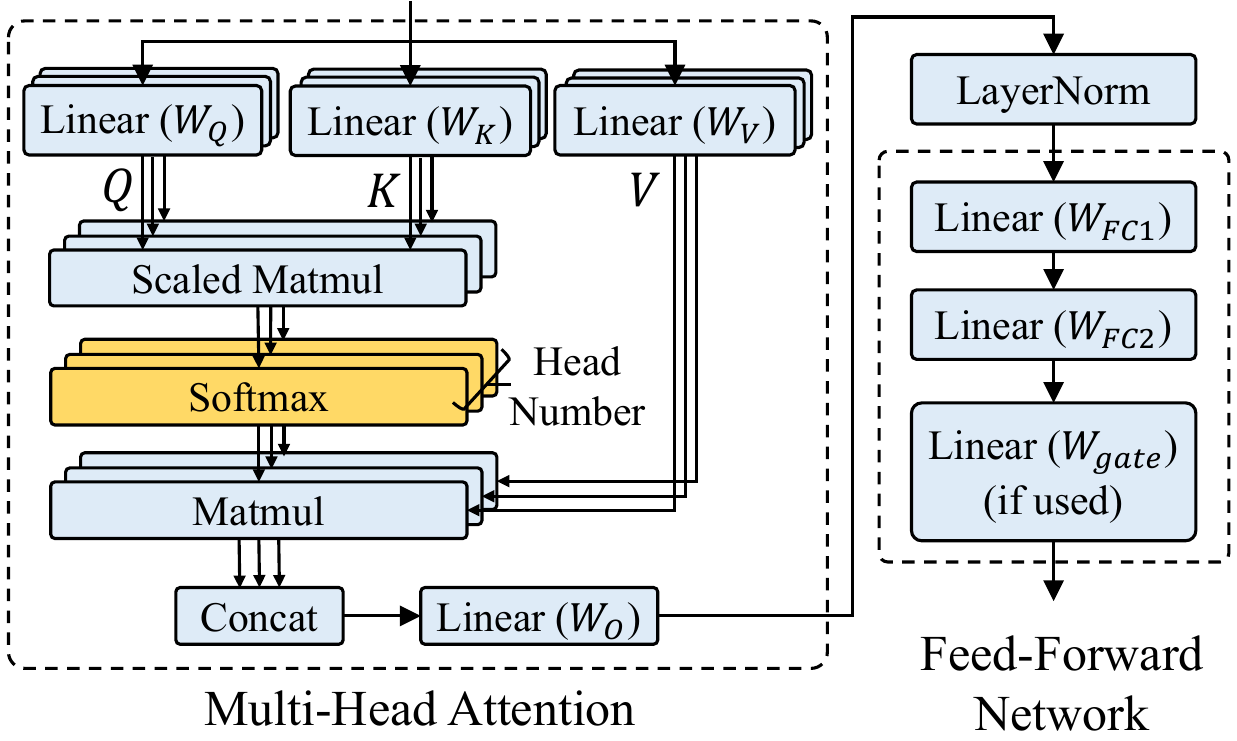}
\caption{Overview of transformer block in Llama2 model.}
\label{LLM}
\end{figure}

While Softmax is used as the final layer only in deep learning works (like CNNs and LSTMs), it is significant for LLMs as it is a part of the attention mechanism as shown in Fig. \ref{LLM}. Previous works have shown that Softmax becomes a bottleneck for LLMs with longer input sequences, attributing to more than 30\% of the overall LLM's latency \cite{stevens2021softermax, liu2024consmax}. We have characterized Softmax for Llama2-7b on A100 GPU and the results are shown in Fig. \ref{runtime_breakdown}. Our characterization reveals that the fraction of time occupied by Softmax increases to 38\% for a sequence length of 16384. Moreover, when GEneral Matrix Multiplication (GEMM)-based operators are accelerated, the non-GEMM operators including Softmax bottleneck the execution time. \cite{karami2024nongemm} reveals that the percentage of execution time occupied by GEMM-operators when performing inference of GPT2-XL on CPU alone was 62\% while that of non-GEMM operators was 38\%. Using a GPU to accelerate, the non-GEMM operators' execution time percentage increases to a drastic 78\%, rendering it the bottleneck.

When it comes to Softmax quantization, we identify two research gaps in the literature: 1-) the absence of an integer-only low-precision approximation of Softmax that does not affect the inference of state-of-the-art LLMs and 2-) the absence of a corresponding efficient integer-only custom hardware capable of implementing the low-precision Softmax for different precisions. To that end, we propose in this paper SoftmAP, a software-hardware co-design that enables accelerating integer-only low-precision Softmax on Associative Processors (APs). To bridge the first gap, we use an integer-only approximation for Softmax, and we explore the effect of using low-precision Softmax in the context of State-of-the-Art (SOTA) Llama models. As for the second gap, we propose the In-Memory Compute (IMC)-based APs as an efficient custom hardware capable of supporting a mix of integer-only low precisions. APs are SIMD-like architectures that perform bit-serial, word parallel arithmetic and logical operations \cite{krikelis1994associative,fouda2022memory}. We have chosen the AP as our custom hardware for three reasons: 1-) their IMC nature, whereby the memory wall problem is not an issue, 2-) the fact that Softmax should be applied to all words in a vocabulary in a parallel fashion hence making the AP a good candidate for acceleration, and 3-) the ability of APs to support mixed-precisions due to its bit-serial word-parallel operation. Our contributions are as follows.

\noindent\textbf{1-} To the best of our knowledge, this is the first work that performs a precision sensitivity analysis of an integer-only Softmax approximation to determine which mixed low-precision implementation does not sacrifice the perplexity for LLMs like Llama2 family. 

\noindent\textbf{2-} We propose a mapping to accelerate the approximated integer-only Softmax with the best mixed-precision on the AP.

\noindent\textbf{3-} We characterize the energy and latency of deploying the integer-only approximated Softmax on the AP, RTX3090 GPU, and A100 GPU for Llama2-7b, Llama2-13b, and Llama2-70b. Our analysis reveals that the AP achieves up to $1300\times$ and $12.58\times$ reduction in energy and latency respectively when compared to the GPU counterparts.

The rest is organized as: Section 2 provides background. Section 3 elaborates on our software-hardware co-design methodology. Section 4 presents the experimental setup, and section 5 presents the results. Section 6 concludes the work.

\section{Background}
\subsection{Large Language Models}

The typical backbone of LLMs is a transformer, and the decoder-only transformer architecture is particularly well-suited for generating text sequentially~\cite{llama2,zhang2022opt,brown2020language,chowdhery2023palm,reid2024gemini}.
By leveraging attention mechanisms and deep neural networks, decoder-only transformers can produce high-quality text outputs that exhibit coherence and relevance to the given input.
The transformer consists of input embeddings and several decoder layers. 
In the input embedding process, the input text is tokenized and embedded into dense vector representations.
The decoder layers enable the model to generate output tokens one at a time, conditioned on the previously generated tokens.
In decoder layers, each decoder block includes multi-head attention, layer normalization, and a feed-forward network  (shown in Fig. \ref{LLM}).
The attention mechanisms allow the model to weigh the importance of different words or tokens in the input text when generating output, which enables the model to capture long-range dependencies and contextual information effectively.
In multi-head attention, the input first passes through the multiple linear layers, resulting in multiple queries, keys, and values ($Q,K,V$). Then, each $Q$ slice and $K$ slice are multiplied and scaled to obtain the attention scores (each slice from an attention head).
Softmax is then applied to normalize them, converting them into a probability distribution representing the importance of each token. This step ensures that the model assigns appropriate weights to different tokens, facilitating effective information aggregation and representation learning. The Softmax is performed as: $Softmax(v)_i={e^{v_i}}/{\sum_i e^{v_i}}$
where $v_i$ is the $i^{th}$ word in a dictionary. After Softmax, each slice of attention matrix is multiplied by $V$ slice and the results from all heads are concatenated.
Feed-Forward Neural Networks then process the output of the self-attention and generate representations that capture the relationships between different tokens in the input text.
Layer norm is used to stabilize the training process.

\begin{figure}[t!]
\centering
\includegraphics[width=1\columnwidth]{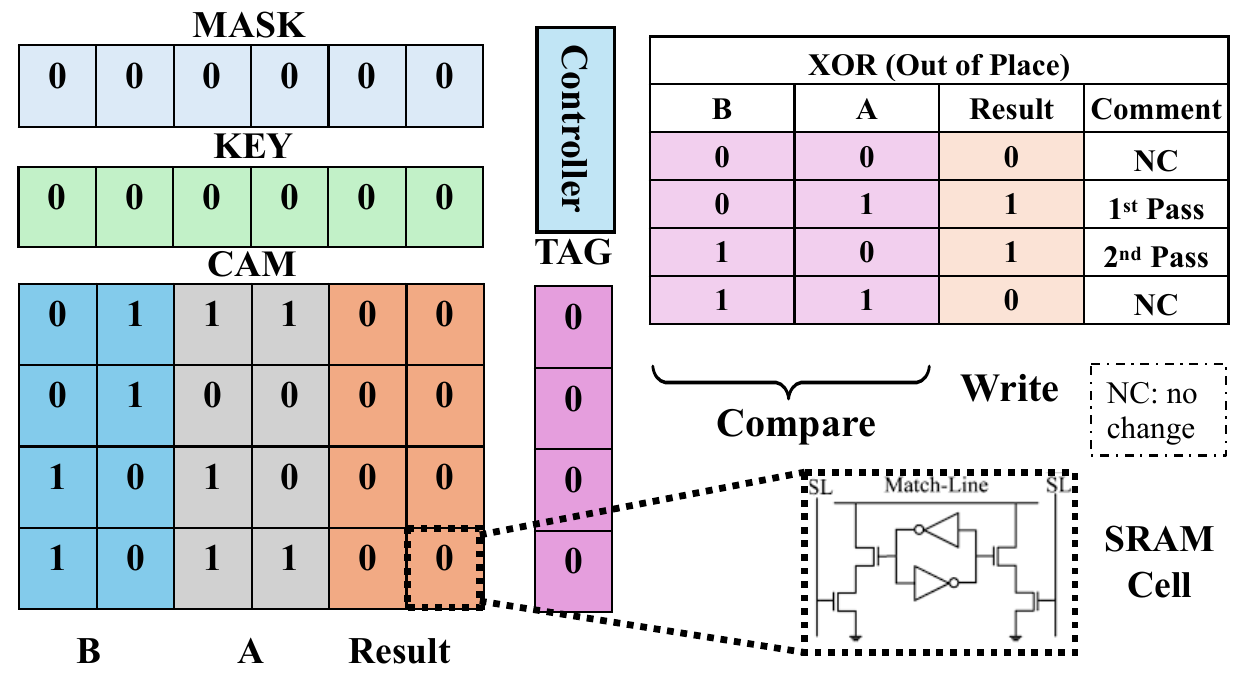}
\caption{SRAM-based AP performing XOR operation between vectors A and B, containing words of precision 2.}
\label{AP}
\end{figure}

\begin{figure*}[b]
\vspace{-0.15in}
\centering
\includegraphics[width=1\textwidth]{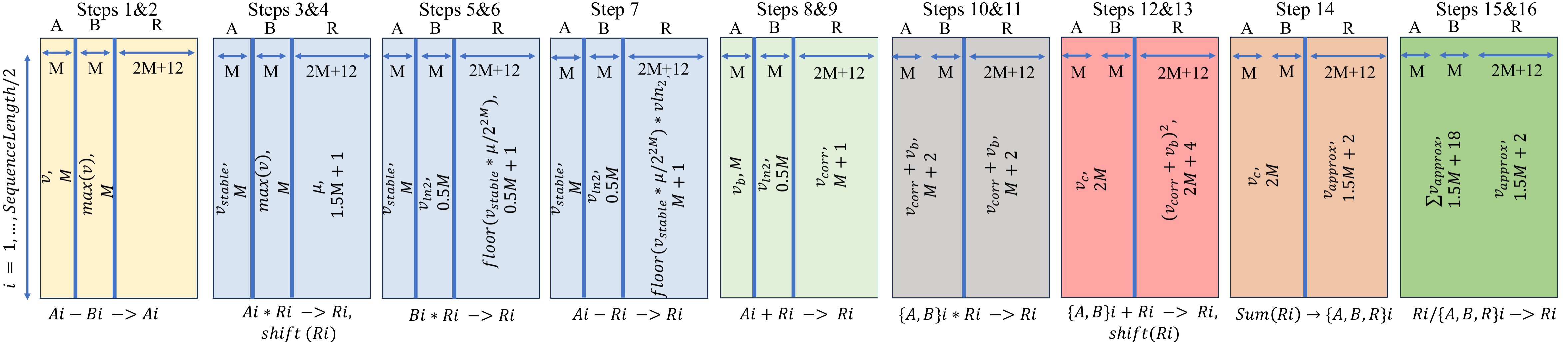}
\caption{Approximate Softmax mapping on one AP inside one head.}
\label{ap_mapping}
\end{figure*}

\subsection{Associative Processors}
An AP's architecture is shown in Fig. \ref{AP}. The building block of the AP is a Content Addressable Memory (CAM), used to store data. The CAM is made up of Static Random Access Memory (SRAM) cells (shown in Fig. \ref{AP}), each storing a bit. The key register is used to present data to be searched in the CAM, and the mask register is used to deactivate certain columns during the search operation if needed. Rows in the CAM that match the presented data are indicated by the tag register. Finally, the controller orchestrates performing the arithmetic/logic operations through a series of compare/write cycles that follow the Look Up Table (LUT) of the corresponding operation. Without loss of generality, fig. \ref{AP} shows an AP storing two vectors A and B, where A=[b'11, b'00, b'10, b'11] and B=[b'01, b'01, b'10, b'10]. The goal is to perform the XOR operation on the adjacent words in the AP by relying on the LUT (Fig. \ref{AP}). First, A and B are written in the AP. Then, the Least Significant Bits (LSBs) are selected with the help of the mask register, and the data of the first pass in the XOR LUT is searched across the AP's CAM rows simultaneously, and matching rows are tagged. After that, the LSBs of the result of matching rows are written according to the 1st pass in the LUT. The 2nd pass comprising a pair of compare/write is then applied to the LSBs according to the LUT. After that, the same compare/write phases are applied to the MSBs according to LUT passes. compare/write on the Most Significant Bits (MSBs) according to the 1st pass of the LUT. Finally, the AP's result=[b'10', b'01', b'00', b'01'] will hold A XOR B. Similar to XOR, any other arithmetic/logic operation can be performed on the AP using the corresponding LUT. For more information about the AP, please refer to \cite{yantir2018efficient}. Works in literature have proposed a two-dimensional AP (2D AP) whereby the same operation that was performed in a bit-serial word-parallel fashion can be applied in the other dimension: selecting two rows at a time and applying the operation in parallel across all their bits. The 2D-AP has an additional set of tag, key, and mask registers. In this work, we rely on the 2D AP as it has reduced complexities compared to the 1D AP, especially for operations amongst rows, like reduction which can be performed without any data movements \cite{yantir2018two}. 

\begin{figure}[H]
\centering
\includegraphics[width=0.9\columnwidth]{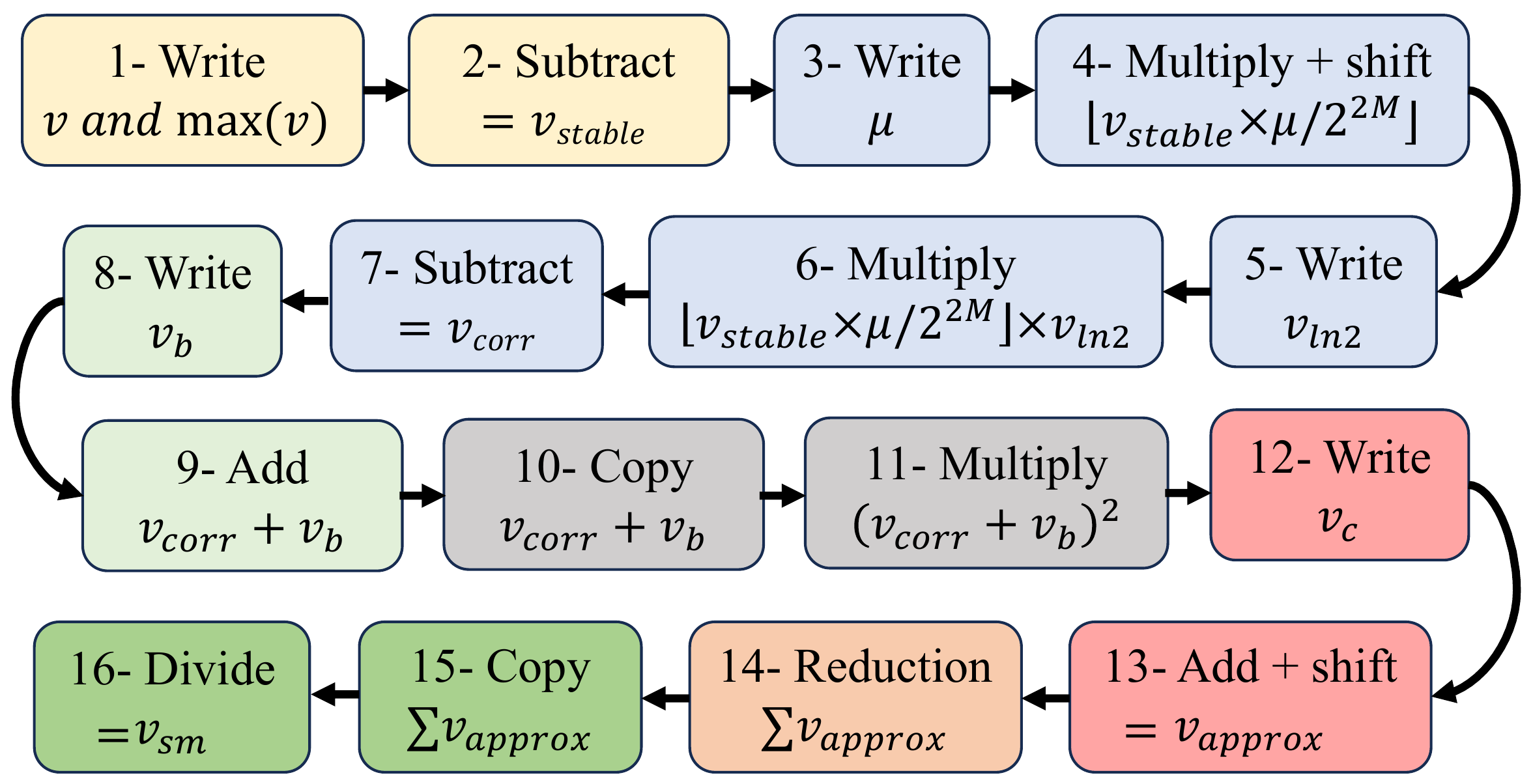}
\caption{AP data flow of the approximate Softmax.}
\label{dataflow}
\end{figure}

\begin{algorithm}[t!]
\caption{Integer-only Softmax approximation}
\label{myalgo}
\begin{algorithmic}[1]
\STATE \textbf{Input:} $v, S, M$
\STATE \textbf{Output:} $v_{sm}, S_{sm}$
\STATE \textbf{Function} \textit{IntOnlySoftmax($v, S$)}
\STATE \hspace {5mm}$v_{stable} \leftarrow v - \max(v)$
\STATE \hspace {5mm}$v_{ln2} \leftarrow \lfloor \ln(2)/S \rfloor$
\STATE \hspace {5mm}{$\mu \leftarrow \lfloor 2^{2M} / v_{ln2} \rfloor$}
\STATE \hspace {5mm}{$v_{corr} \leftarrow v_{stable} - \lfloor v_{stable} \times \mu / 2^{2M} \rfloor \times v_{ln2}$}
\STATE \hspace {5mm}$a, b, c \leftarrow 0.3585, 1.353, 0.344$
\STATE \hspace {5mm}$v_b \leftarrow \lfloor b/S \rfloor$
\STATE \hspace {5mm}$v_c \leftarrow \lfloor c/aS^2 \rfloor$
\STATE \hspace {5mm}$v_{approx} \leftarrow ((v_{corr}+v_b)^2+v_c)>> \lfloor -v_{stable}/v_{ln2}\rfloor$
\STATE \hspace {5mm}$v_{sm}=v_{approx}/\sum{v_{approx}}$
\STATE \hspace {5mm} $S_{sm} \leftarrow \lfloor aS^2 \rfloor$  
\STATE \hspace {5mm} \textbf{return }$v_{sm}, S_{sm}$
\end{algorithmic}
\end{algorithm}

\section{SoftmAP}
\textbf{Softmax Approximation:} To approximate Softmax, we use integer polynomial approximations and Barret reduction \cite{barrett1986implementing,kim2021bert} to speed up its deployment on our hardware. The approximation is shown in Algorithm \ref{myalgo}, which approximates the exponential function in Softmax by a second-order polynomial, hence involving integer-only computations. The algorithm takes the $M$-bit quantized values in a vocabulary $v_i$ (we omit henceforth the "i" for brevity) and their scaling factor $S$. Since we limit the scaling factor according to the range of data in the given dataset, this number can be computed offline. The first step in the algorithm (line 4) is to subtract the maximum quantized value from the quantized input for numerical stability, as this would avoid underflow/overflow caused by the exponent \cite{milakov2018online}. Since $v_{stable}$ is a non-positive number, it is decomposed into a non-negative integer quotient $q$ and a real remainder $r$ $\in (-ln(2),0]$. It then follows that $e^{v_{stable}}$ can be approximated by $e^r$ right shifted by $q$ as shown in \cite{kim2021bert}. Kim et al. show that $e^r$ can be approximated using second-order polynomial and integer computations. The second order polynomial is $(v_{corr}+v_b)^2+v_c$ where $v_{corr}=r=(v_{stable} \quad mod \quad v_{ln2}$) (line 11). $v_{ln2}$ and the coefficients used in the second-order polynomial are computed in lines 5 and 8-10. Since S is fixed, these values can be computed offline. Using the Barret reduction, the computations of the modulus operation $v_{corr}$ (line 7) only involve multiplications, subtractions, and shifts (faster operations than division). This is facilitated by a value $\mu$ precomputed offline (line 6). Finally, the integer-only Softmax is computed as follows: $IntSoftmax(v)=v_{sm}*S_{sm}$ where $v_{sm}$ and $S_{sm}$ are computed in lines 12-13.

\noindent \textbf{Hardware-Friendly Deployment on AP:} To accelerate the approximated Softmax, we map the dataflow presented in Fig. \ref{dataflow} to the 2D AP as shown in Fig. \ref{ap_mapping}, where this AP is deployed in each head. We note that we use the same color code in both figures. For each computation, the AP applies the LUT of the corresponding operation, in parallel, across $SequenceLength/2$ words. Precisions shown in Fig. \ref{ap_mapping} are for $M=8$ $v_{corr}=M$, and $N=16$ in Table \ref{precisions}. The result column "R" can fit $2M+12$ bits (this big precision is required in the last step to store the final result in this column, after the division is performed), but in some steps, we store lower precision words in this column (for example Steps 3 and 4). Steps 1 and 2 of the data flow are represented in the first AP illustration whereby columns $A$ and $B$ hold each $SequenceLength/2$ words $v$ and $max(v)$, each of precision $M$. $v-max(v)$ is computed in parallel across the rows of the AP, bit-serially, and following the corresponding LUT. The resulting vector $v_{stable}$ (of precision $M$) overwrites the rows of column $A$, as shown vertically in the second illustration. A similar approach is followed to complete the Softmax computation (steps 3 to 13) in one head using the same AP. We note that in some cases, we need to store words in more than one column (steps 10 to 13), in which case we represent the computation inside braces. For example $\{A,B\}i+Ri -> Ri$, this means we are adding words $Ri$ written in column R to the words written in columns A and parts of column B, $\{A,B\}i$. Since $\mu$, $a$, $b$, $c$, $v_b$, $v_c$, and $v_{ln2}$ are computed offline, we only need to write them into the AP during execution. Shift operations are inherently supported by bit-seriality of the AP.

\begin{table}[t!]
\caption{Precisions used for software implementation. \textbf{sum}\boldmath{$=\sum v_{approx}$} and $N=$ additional bits needed to store the sum.}
\label{precisions}
\resizebox{\columnwidth}{!}{
\begin{tabular}{ccccc|ccc|ccc|}
\Xcline{3-11}{1.0pt}
 &
   &
  \multicolumn{3}{[c]}{\boldmath{$v_{corr}=M$}} &
  \multicolumn{3}{c]}{\boldmath{$v_{corr}=M+1$}} &
  \multicolumn{3}{c]}{\boldmath{$v_{corr}=M+2$}} \\ \thickhline
\multicolumn{2}{[c}{\boldmath{$M$}} &
  \multicolumn{1}{[c|}{\textbf{4}} &
  \multicolumn{1}{c|}{\textbf{6}} &
  \multicolumn{1}{c]}{\textbf{8}} &
  \multicolumn{1}{c|}{\textbf{4}} &
  \multicolumn{1}{c|}{\textbf{6}} &
  \multicolumn{1}{c]}{\textbf{8}} &
  \multicolumn{1}{c|}{\textbf{4}} &
  \multicolumn{1}{c|}{\textbf{6}} &
  \multicolumn{1}{c]}{\textbf{8}} \\ \thickhline
\multicolumn{2}{[c}{\boldmath{$v$}} &
  \multicolumn{1}{[c|}{4} &
  \multicolumn{1}{c|}{6} &
  \multicolumn{1}{c]}{8} &
  \multicolumn{1}{c|}{4} &
  \multicolumn{1}{c|}{6} &
  \multicolumn{1}{c]}{8} &
  \multicolumn{1}{c|}{4} &
  \multicolumn{1}{c|}{6} &
  \multicolumn{1}{c]}{8} \\ \hline
\multicolumn{2}{[c}{\boldmath{$v_{stable}$}} &
  \multicolumn{1}{[c|}{4} &
  \multicolumn{1}{c|}{6} &
  \multicolumn{1}{c]}{8} &
  \multicolumn{1}{c|}{4} &
  \multicolumn{1}{c|}{6} &
  \multicolumn{1}{c]}{8} &
  \multicolumn{1}{c|}{4} &
  \multicolumn{1}{c|}{6} &
  \multicolumn{1}{c]}{8} \\ \hline
\multicolumn{2}{[c}{\boldmath{$v_{ln2}$}} &
  \multicolumn{1}{[c|}{4} &
  \multicolumn{1}{c|}{4} &
  \multicolumn{1}{c]}{4} &
  \multicolumn{1}{c|}{4} &
  \multicolumn{1}{c|}{4} &
  \multicolumn{1}{c]}{4} &
  \multicolumn{1}{c|}{4} &
  \multicolumn{1}{c|}{4} &
  \multicolumn{1}{c]}{4} \\ \hline
\multicolumn{2}{[c}{\boldmath{$v_b$}} &
  \multicolumn{1}{[c|}{4} &
  \multicolumn{1}{c|}{6} &
  \multicolumn{1}{c]}{8} &
  \multicolumn{1}{c|}{4} &
  \multicolumn{1}{c|}{6} &
  \multicolumn{1}{c]}{8} &
  \multicolumn{1}{c|}{4} &
  \multicolumn{1}{c|}{6} &
  \multicolumn{1}{c]}{8} \\ \hline
\multicolumn{2}{[c}{\boldmath{$v_c$}} &
  \multicolumn{1}{[c|}{8} &
  \multicolumn{1}{c|}{12} &
  \multicolumn{1}{c]}{16} &
  \multicolumn{1}{c|}{8} &
  \multicolumn{1}{c|}{12} &
  \multicolumn{1}{c]}{16} &
  \multicolumn{1}{c|}{8} &
  \multicolumn{1}{c|}{12} &
  \multicolumn{1}{c]}{16} \\ \hline
\multicolumn{2}{[c}{\boldmath{$(v_{corr}+v_b)^2+v_c$}} &
  \multicolumn{1}{[c|}{11} &
  \multicolumn{1}{c|}{15} &
  \multicolumn{1}{c]}{19} &
  \multicolumn{1}{c|}{13} &
  \multicolumn{1}{c|}{17} &
  \multicolumn{1}{c]}{21} &
  \multicolumn{1}{c|}{15} &
  \multicolumn{1}{c|}{19} &
  \multicolumn{1}{c]}{23} \\ \hline
\multicolumn{2}{[c}{\boldmath{$v_{approx}$}} &
  \multicolumn{1}{[c|}{10} &
  \multicolumn{1}{c|}{12} &
  \multicolumn{1}{c]}{14} &
  \multicolumn{1}{c|}{12} &
  \multicolumn{1}{c|}{14} &
  \multicolumn{1}{c]}{16} &
  \multicolumn{1}{c|}{14} &
  \multicolumn{1}{c|}{16} &
  \multicolumn{1}{c]}{18} \\ \thickhline
\multicolumn{1}{[c|}{\multirow{4}{*}{\textbf{sum}}} &
  \textbf{N=8} &
  \multicolumn{1}{[c|}{18} &
  \multicolumn{1}{c|}{20} &
  \multicolumn{1}{c]}{22} &
  \multicolumn{1}{c|}{20} &
  \multicolumn{1}{c|}{22} &
  \multicolumn{1}{c]}{24} &
  \multicolumn{1}{c|}{22} &
  \multicolumn{1}{c|}{24} &
  \multicolumn{1}{c]}{26} \\ \cline{2-11} 
\multicolumn{1}{[c|}{} &
  \textbf{N=12} &
  \multicolumn{1}{[c|}{22} &
  \multicolumn{1}{c|}{24} &
  \multicolumn{1}{c]}{26} &
  \multicolumn{1}{c|}{24} &
  \multicolumn{1}{c|}{26} &
  \multicolumn{1}{c]}{28} &
  \multicolumn{1}{c|}{26} &
  \multicolumn{1}{c|}{28} &
  \multicolumn{1}{c]}{30} \\ \cline{2-11} 
\multicolumn{1}{[c|}{} &
  \textbf{N=16} &
  \multicolumn{1}{[c|}{26} &
  \multicolumn{1}{c|}{28} &
  \multicolumn{1}{c]}{30} &
  \multicolumn{1}{c|}{28} &
  \multicolumn{1}{c|}{30} &
  \multicolumn{1}{c]}{32} &
  \multicolumn{1}{c|}{30} &
  \multicolumn{1}{c|}{32} &
  \multicolumn{1}{c]}{34} \\ \cline{2-11} 
\multicolumn{1}{[c|}{} &
  \textbf{N=20} &
  \multicolumn{1}{[c|}{30} &
  \multicolumn{1}{c|}{32} &
  \multicolumn{1}{c]}{34} &
  \multicolumn{1}{c|}{32} &
  \multicolumn{1}{c|}{34} &
  \multicolumn{1}{c]}{36} &
  \multicolumn{1}{c|}{34} &
  \multicolumn{1}{c|}{36} &
  \multicolumn{1}{c]}{38} \\ \thickhline
\end{tabular}

}
\vspace{-0.15in}
\end{table}

\section{Experimental Setup}
We integrate the Softmax approximation in the 7B, 13B, and 70B models in the Llama2 family~\cite{llama2}, and vary the precisions according to Table \ref{precisions} to estimate the effect of different precision combinations on the perplexity. Note that $\mathbf{N}$ represents the additional bits needed to store the $\mathbf{sum}=\sum_i e^{v_i}$. In the scenario where we do not truncate the sum, $\mathbf{N}=log_2(\frac{SequenceLength}{2})$. The perplexity is evaluated on WikiText-2 benchmark~\cite{wikitext}. Our implementation uses PyTorch 2.0.1~\cite{pytorch} and Transformers 4.41.2~\cite{transformer}. And all models are obtained from HuggingFace~\cite{HuggingFace}.
For language generation tasks, we calculate the perplexity as follows: (1) The validation set is concatenated using two linebreaks as separators and encoded using the default HuggingFace tokenizer of each model. (2) The sequence is split into non-overlapping segments of width $2048$, the full context size of our models. (3) Then, they are sent to the model to obtain the log probabilities corresponding to the next token, and their exponentiated average is the final standard perplexity. 
We evaluate all individual samples separately and do not apply any padding. We execute the software experiments on one NVIDIA RTX3090 GPU and one NVIDIA A100 GPU.
\begin{table}[t!]
\caption{Run time of functions. $L$ is \#words in AP, $M$ is precision/word, matrices for multiplication are $i\times j$ and $j\times u$.}
\label{runtime}
\centering
\resizebox{0.5\textwidth}{!}{%
\begin{tabular}{|*{2}{c|}}
\hline
\textbf{Function}& \multicolumn{1}{|c|}{\textbf{2D AP runtime}} \\ \hline
Addition & \multicolumn{1}{|c|}{$2M+8M+M+1$}\\
\hline
Multiplication & \multicolumn{1}{|c|}{$2M+8M^2+2M$}\\
\hline
Reduction & \multicolumn{1}{|c|}{$2M+8M+8\log_2(L/2)+1$}\\
\hline
Matrix-Matrix Multiplication & \multicolumn{1}{|c|}{$2M+8M^2+8\log_2(j)+2M+\log_2(j)$}\\ \hline
\end{tabular}
}
\vspace{-0.2in}
\end{table}

For the custom hardware cost estimation, we have designed a Python-based AP simulator that models the data flow execution of the integer-only Softmax approximation shown in Fig. \ref{dataflow}. Our simulator models the SRAM-based AP assuming a $16nm$ technology and relies on the formulations in Table \ref{runtime} to model the energy and latency of performing elementary operations (addition, multiplication, etc.) similar to \cite{rakka2024bf}. Using the elementary operations and the dataflow, we find the overall energy/latency required to perform the Softmax. 

 \begin{table}[h!]
 \vspace{0.15in}
 \caption{Perplexity for Llama2-7b with input in [-7,0].}
\label{perplexity1}
\resizebox{\columnwidth}{!}{
\begin{tabular}{|c|cc|cc|cc|}
\hline
\multirow{2}{*}{\textbf{\begin{tabular}[c]{@{}c@{}}Perplexity($\downarrow$)\\ Llama2-7b\end{tabular}}} &
  \multicolumn{2}{c|}{\boldmath{$v_{corr}=M$}} &
  \multicolumn{2}{c|}{\boldmath{$v_{corr}=M+1$}} &
  \multicolumn{2}{c|}{\boldmath{$v_{corr}=M+2$}} \\ \cline{2-7} 
 &
  
  \multicolumn{1}{c|}{\textbf{M=6}} &
  \textbf{M=8} &
  
  \multicolumn{1}{c|}{\textbf{M=6}} &
  \textbf{M=8} &
  
  \multicolumn{1}{c|}{\textbf{M=6}} &
  \textbf{M=8} \\ \hline
\textbf{N=8} &
 
  \multicolumn{1}{c|}{9.62} &
  17.78 &

  \multicolumn{1}{c|}{9.62} &
  17.77 &

  \multicolumn{1}{c|}{9.62} &
  17.77 \\ \hline
\textbf{N=12} &

  \multicolumn{1}{c|}{5.92} &
  5.52 &

  \multicolumn{1}{c|}{5.93} &
  5.52 &

  \multicolumn{1}{c|}{5.93} &
  5.52 \\ \hline
\textbf{N=16} &

  \multicolumn{1}{c|}{5.92} &
  {\underline{\textbf{5.51}}} &

  \multicolumn{1}{c|}{5.92} &
  5.51 &

  \multicolumn{1}{c|}{5.92} &
  5.51 \\ \hline
\textbf{N=20} &

  \multicolumn{1}{c|}{5.92} &
  5.51 &

  \multicolumn{1}{c|}{5.92} &
  5.51 &

  \multicolumn{1}{c|}{5.92} &
  5.51 \\ \hline
  \multicolumn{7}{|c|}{\textbf{FP Perplexity = 5.47}} \\ \hline
\end{tabular}
}
\end{table}
\begin{table}[h!]
\caption{Perplexity for Llama2-13b with input in [-7,0].}
\label{perplexity3}
\resizebox{\columnwidth}{!}{
\begin{tabular}{|c|cc|cc|cc|}
\hline
\multirow{2}{*}{\textbf{\begin{tabular}[c]{@{}c@{}}Perplexity($\downarrow$)\\ Llama2-13b\end{tabular}}} &
  \multicolumn{2}{c|}{\boldmath{$v_{corr}=M$}} &
  \multicolumn{2}{c|}{\boldmath{$v_{corr}=M+1$}} &
  \multicolumn{2}{c|}{\boldmath{$v_{corr}=M+2$}} \\ \cline{2-7} 
 &
  
  \multicolumn{1}{c|}{\textbf{M=6}} &
  \textbf{M=8} &
  \multicolumn{1}{c|}{\textbf{M=6}} &
  \textbf{M=8} &
  \multicolumn{1}{c|}{\textbf{M=6}} &
  \textbf{M=8} \\ \hline
\textbf{N=8}  & \multicolumn{1}{c|}{13.38} & 12.78 & \multicolumn{1}{c|}{13.38} & 12.8 & \multicolumn{1}{c|}{13.38} & 12.78 \\ \hline
\textbf{N=12} & \multicolumn{1}{c|}{5.54}  & 4.94  & \multicolumn{1}{c|}{5.54}  & 4.94 & \multicolumn{1}{c|}{5.54}  & 4.94  \\ \hline
\textbf{N=16} & \multicolumn{1}{c|}{5.35}  & {\underline{\textbf{4.93}}}  & \multicolumn{1}{c|}{5.35}  & 4.93 & \multicolumn{1}{c|}{5.35}  & 4.93  \\ \hline
\textbf{N=20} & \multicolumn{1}{c|}{5.34}  & 4.93  & \multicolumn{1}{c|}{5.34}  & 4.93 & \multicolumn{1}{c|}{5.34}  & 4.93  \\ \hline
\multicolumn{7}{|c|}{\textbf{FP Perplexity = 4.88}} \\ \hline
\end{tabular}
}
\vspace{-0.15in}
\end{table}

\begin{figure*}[!h]
\centering
\subfloat[]{\includegraphics[width=0.333\textwidth]
{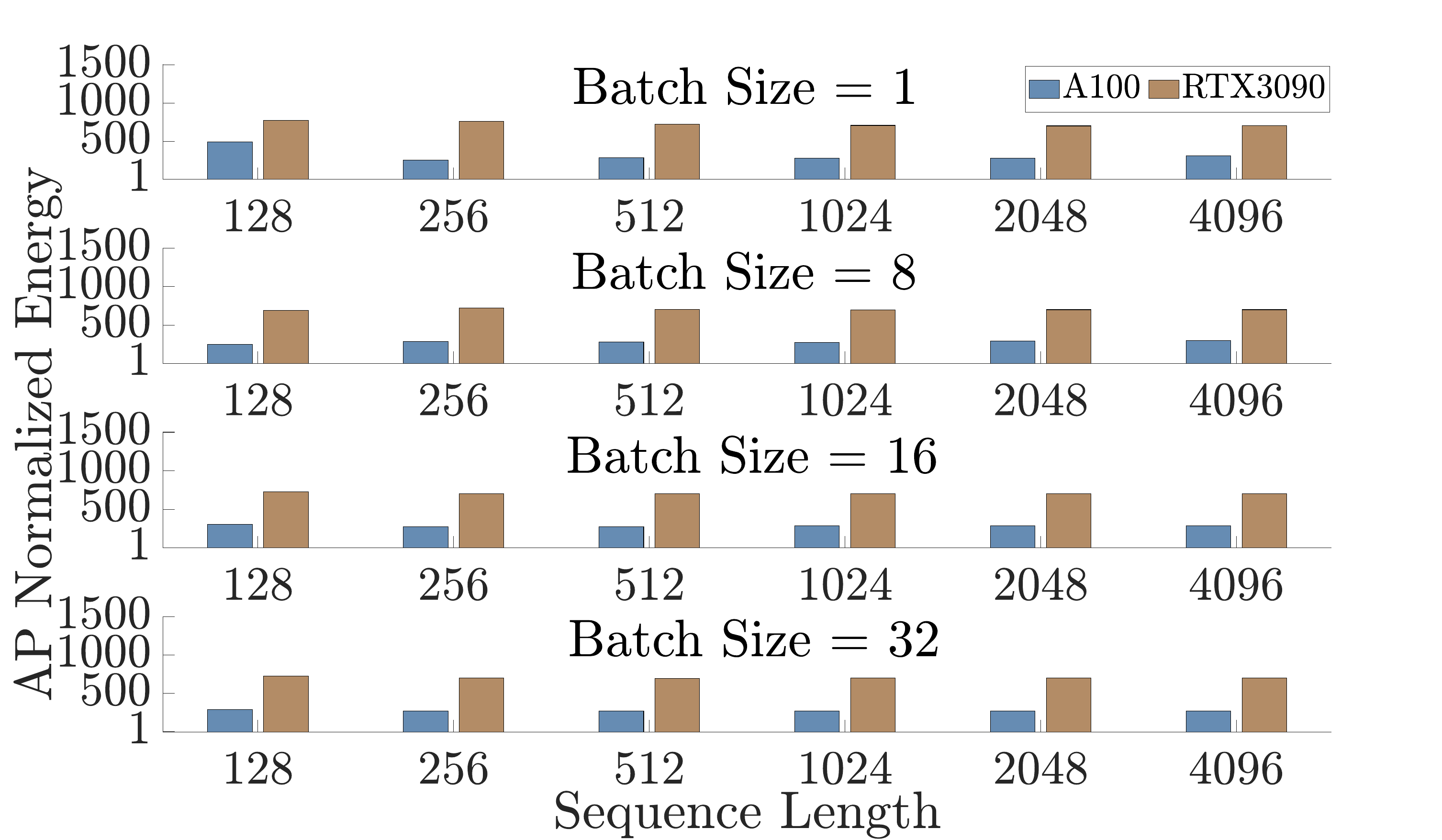}}
\hfil
\subfloat[]
{\includegraphics[width=0.333\textwidth]
{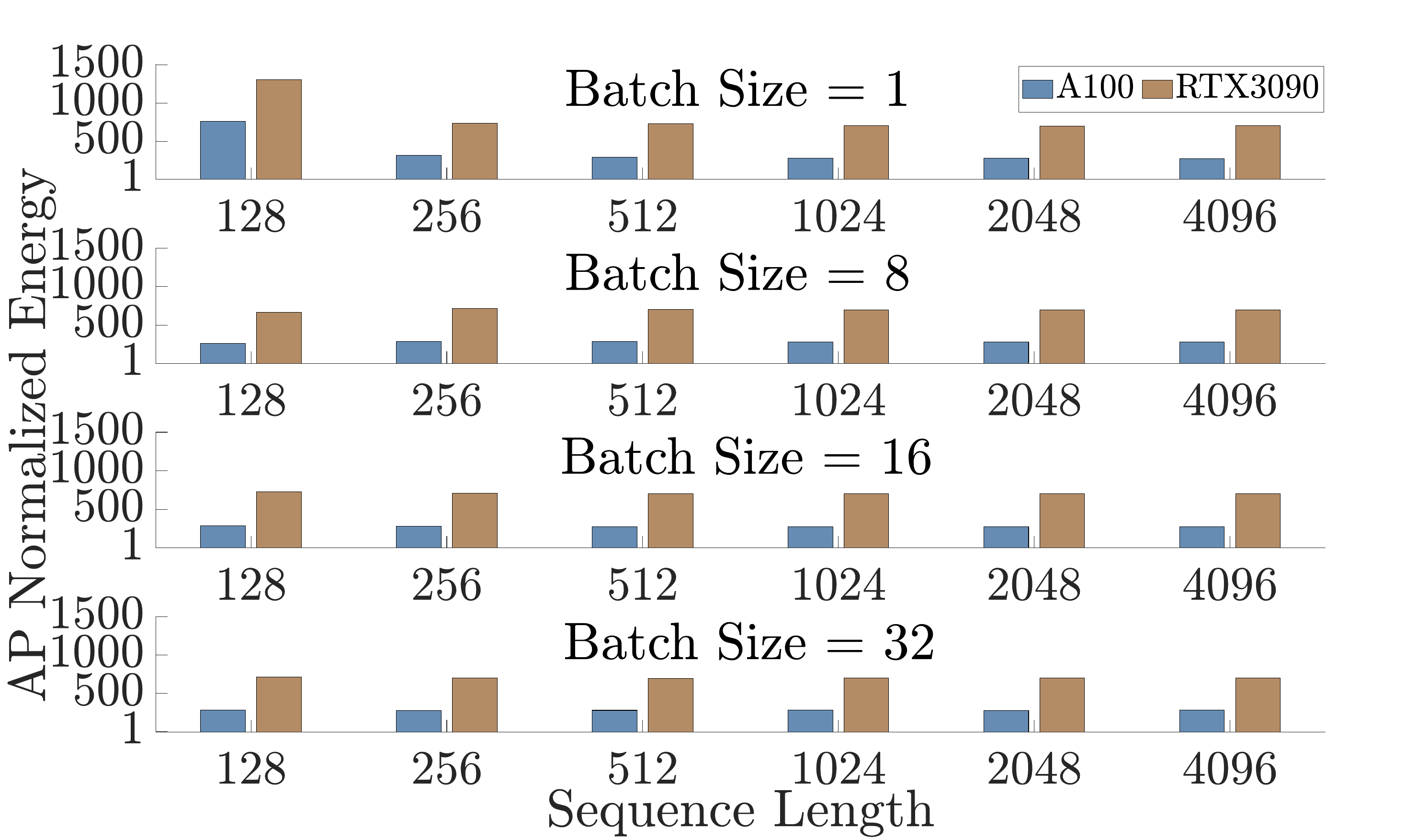}}
\hfil
\subfloat[]
{\includegraphics[width=0.333\textwidth]
{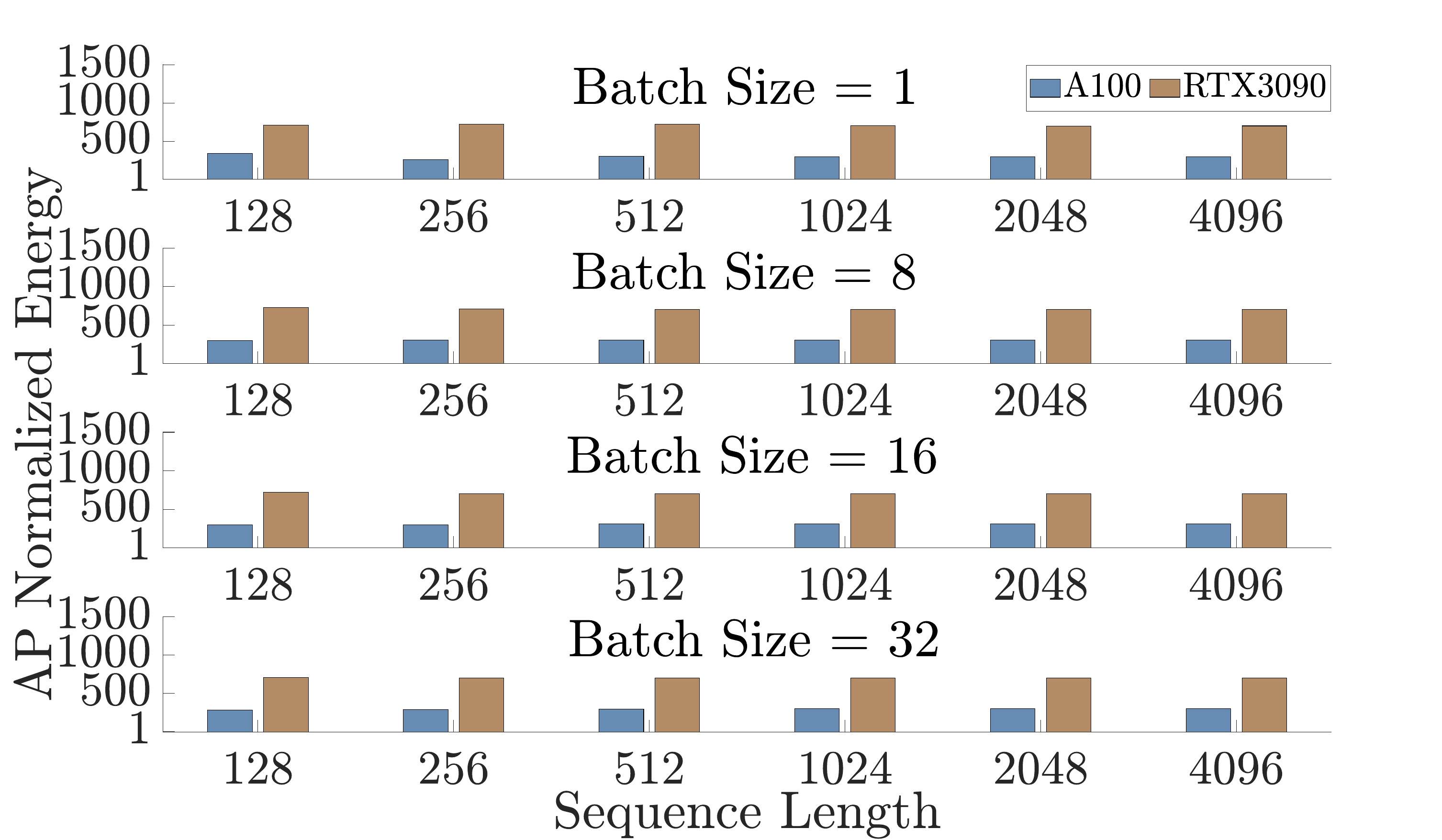}}
\caption{AP normalized energy for (a) Llama2-7b, (b) Llama2-13b, and (c) Llama2-70b.}
\label{energy}
\vspace{-0.1in}
\end{figure*}

\begin{figure*}[!h]
\centering
\subfloat[]{\includegraphics[width=0.333\textwidth]
{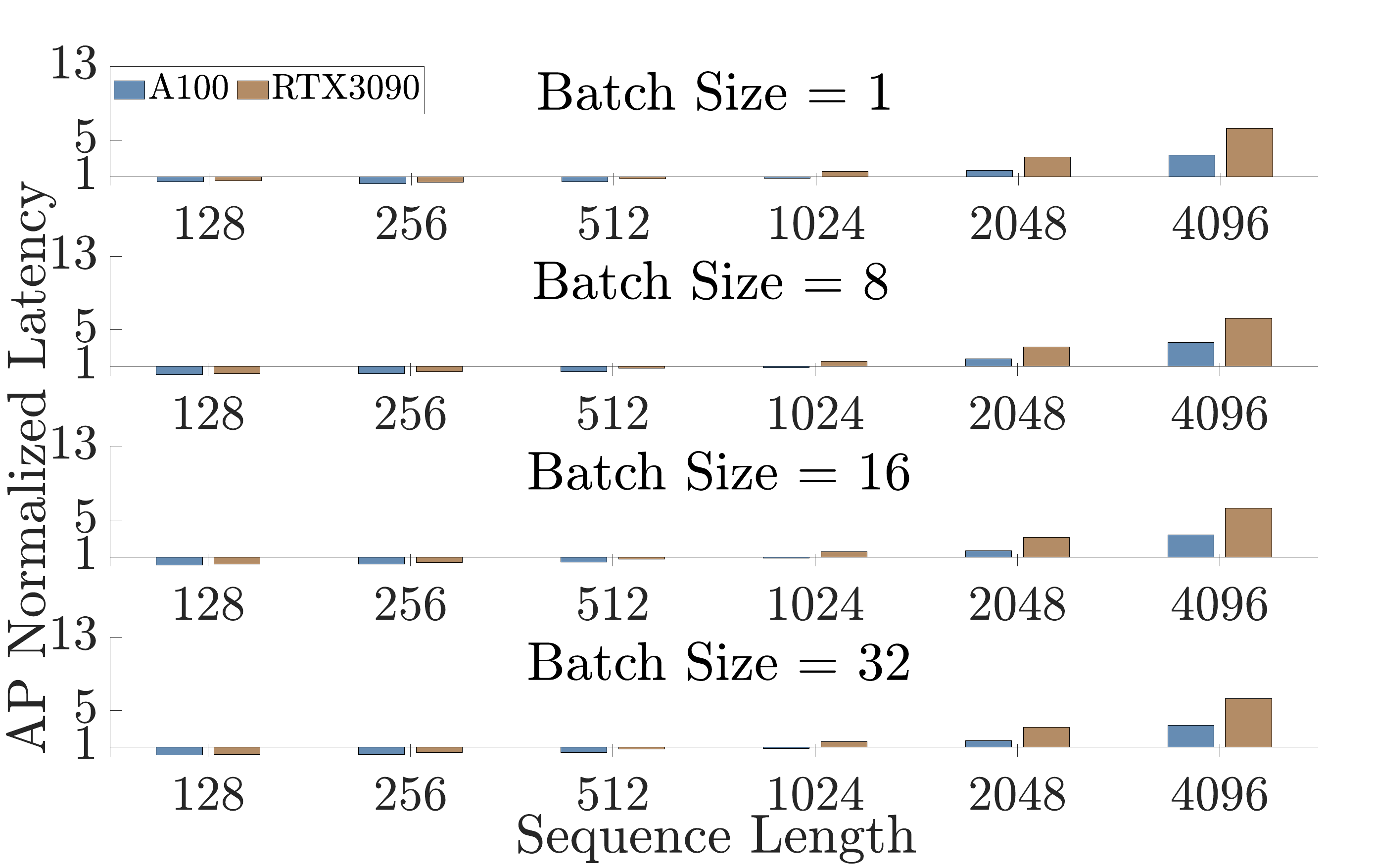}}
\hfil
\subfloat[]
{\includegraphics[width=0.333\textwidth]
{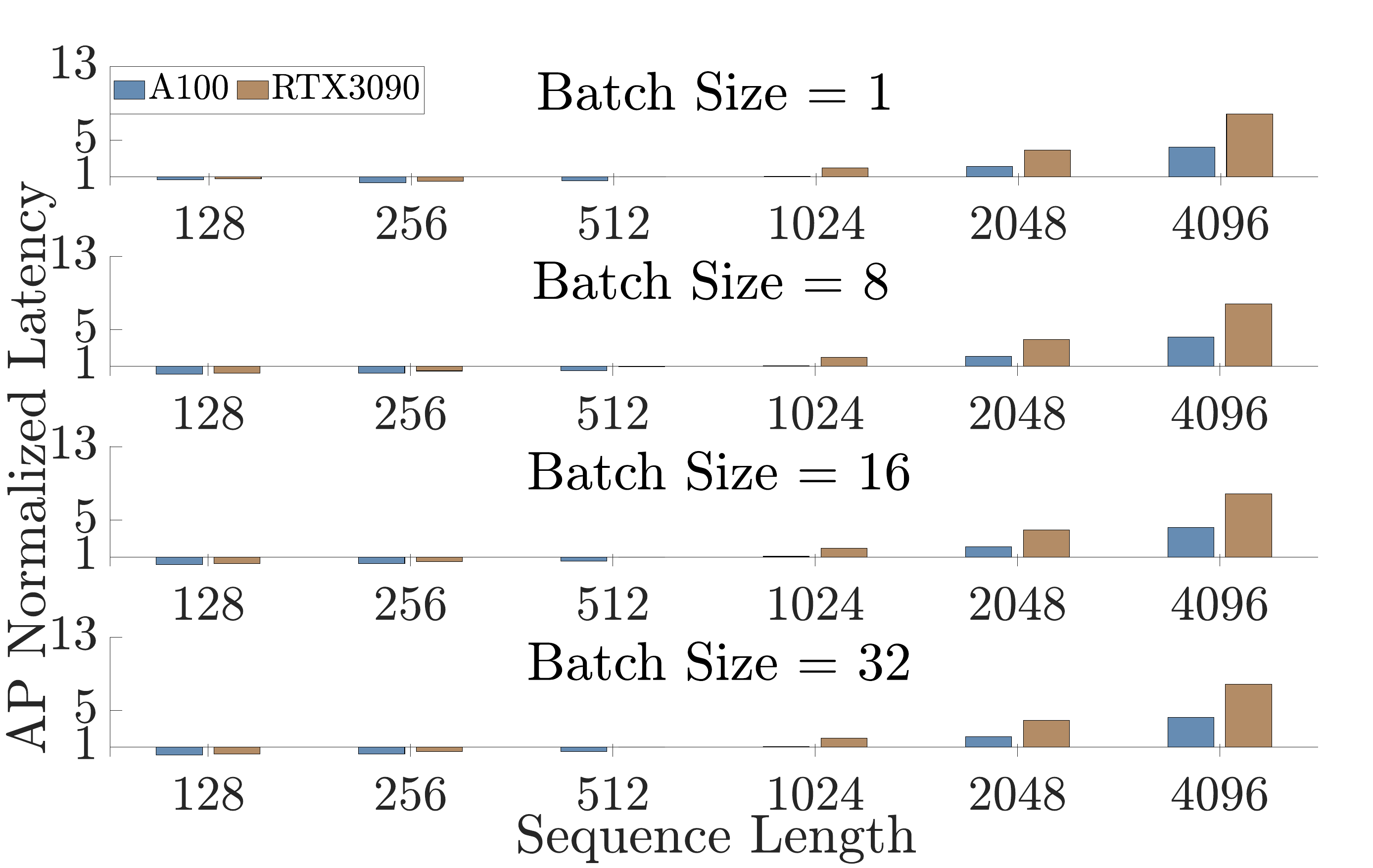}}
\hfil
\subfloat[]
{\includegraphics[width=0.333\textwidth]
{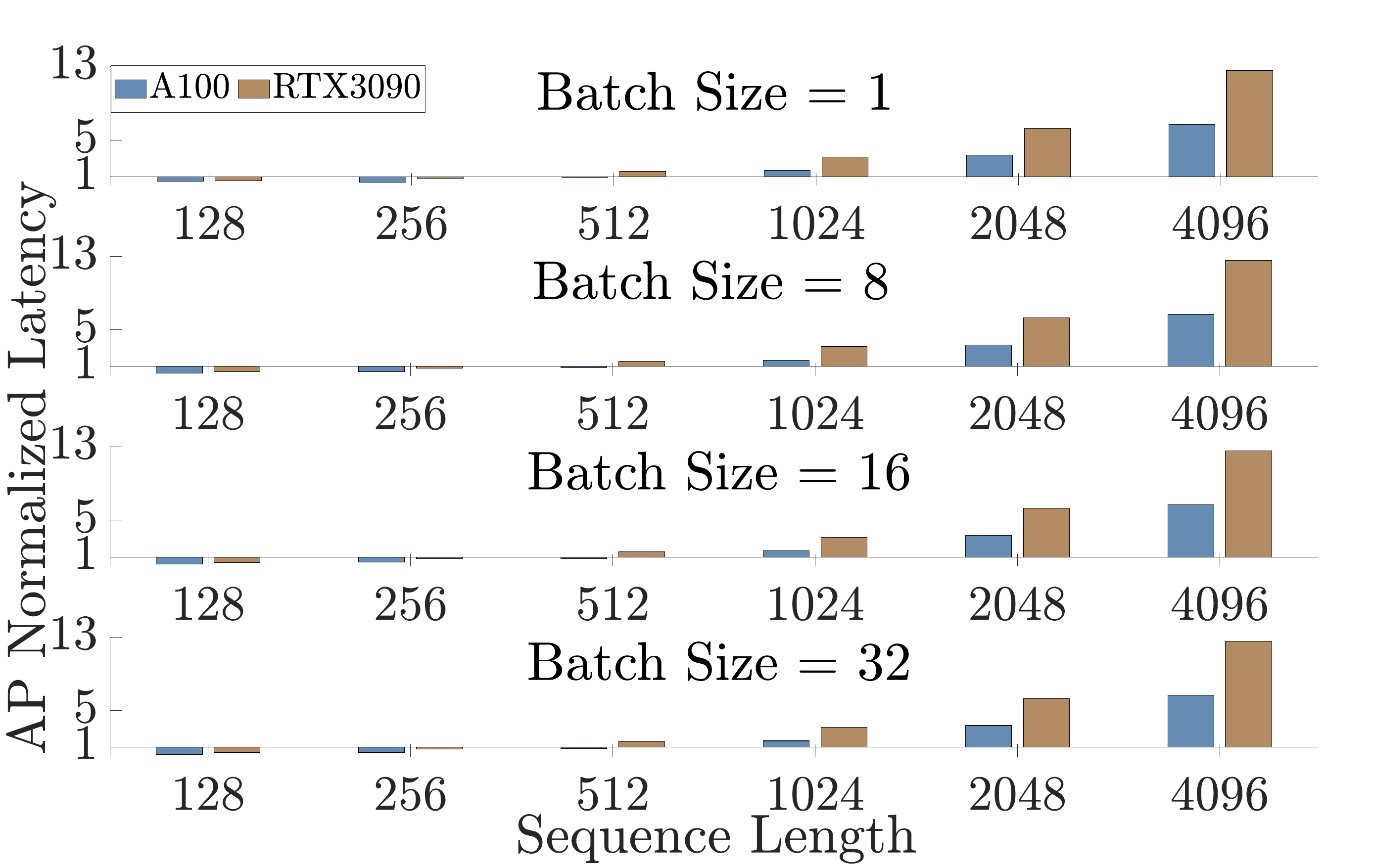}}
\caption{AP normalized latency for (a) Llama2-7b, (b) Llama2-13b, and (c) Llama2-70b.}
\label{latency}
\end{figure*}

\section{Results and Discussion}
\subsection{Precision Sensitivity Analysis}
First, we evaluate the sensitivity of our approximated integer-only Softmax to precision. We study the perplexity when using the different mixed-precision combinations reported in Table \ref{precisions} for the different parts of the approximated Softmax Algorithm~\ref{myalgo}. We present in Table \ref{perplexity1} the perplexity values of Llama2-7b when performing end-to-end inference using the approximated integer-only Softmax and with input clipping to $[T_C,0]$, where $T_C=-7$. 
The selection of clipping threshold \( T_C \) primarily considers two factors: a larger \( T_C \) increases the quantization error within the \([T_C, 0]\) range of the Softmax input, while a smaller \( T_C \) reduces the quantization error in this range but causes significant clipping in the \((-\infty, T_C)\) range, leading to dominant precision loss. By using the WikiText-2 dataset as a calibration dataset, we analyze the input range of Softmax and manually select \( T_C=-7 \) with $M\in\{6,8\}$ and \( T_C=-4 \) with $M=4$ as global clipping parameters for the model. We believe that a more refined selection process could further enhance quantization precision.
We omit the results for $M=4$ from Table \ref{perplexity1} for the following reason: When $M=4$, we set the clipping threshold to $-4$ (the quantized values have larger absolute differences, so the clipping threshold needs to be closer to zero to ensure the resolution), the resulting perplexities values were between $176$ and $46$, i.e. $8\times$-$32\times$ worse than the FP Softmax perplexity (which is $5.47$). This indicates that $M=4$ is too small to yield a good perplexity and must not be used. From Table \ref{perplexity1}, we notice that varying the precision of $v_{corr}$ does not affect perplexity at all and that for a given $M$, as we increase the value of $N$, the perplexity decreases up to $N=16$, after which no change in perplexity is reported. In addition, we see that $M=8$ yields better perplexity than $M=6$ for a given $N$ and $v_{corr}$. A similar analysis can be made for Llama2-13b and Llama2-70b. Due to space limitations, we only report the perplexities of Llama2-13b in Table \ref{perplexity3} for input clipping of $[-7,0]$. By observing the perplexity values, we choose $v_{corr}=M$, $M=6$, and $N=16$ as the ``best precision combination" that yields the lowest perplexity with the lowest bitwidths all three Llama models. Next, we use this precision combination to analyze the energy and latency. 

\vspace{-0.05in}
\subsection{Hardware Evaluation}
We characterize the approximated Softmax with the best precision combination using Llama2-7b, Llama2-13b, and Llama2-70b on the GPU platforms (A100 and RTX3090) and our proposed AP simulator, and we present the normalized (with respect to the AP) energy and latency values in Figs. \ref{energy} and \ref{latency}. Each normalized energy value reported is defined as ${Energy_{GPU}}/{Energy_{AP}}$. The normalized latency is similarly defined. Figs. \ref{energy} (a), (b), and (c) show that the AP is more energy efficient than the A100 and RTX3090 GPUs for the Llama models for all sequence lengths and all batch sizes. In fact, the AP reports up to $489\times$($760\times$)($340\times$) and $776\times$($1305\times$)($726\times$) energy savings compared to A100 and RTX3090 with on average savings being $289\times$($301\times$)($301\times$) and $710\times$($730\times$)($707\times$) respectively for Llama2-7b(Llama2-13b)(Llama2-70b). Notably, the highest energy savings for the AP compared with A100 and RTX3090 are reported for a batch size of 1 and a sequence length of 128. For this sequence length and batch size, the GPUs are least efficient, exhibiting much higher energy compared to the AP. As the sequence length and batch size increase, the gap between the GPUs and AP decreases, hence the ratio remains almost constant.

Normalized latency ratios are shown in Fig. \ref{latency}. A ratio above $1$ favors the AP. For Llama2-7b and Llama2-13b and sequence lengths below $1024$ (i.e. $128$, $256$, and $512$), the AP needs more time than the GPUs. For Llama2-70b, this pattern is observed for a sequence length below $512$. This is due to 1-) the bit-serial, word-parallel operation of the AP and 2-) the fact that sequence length affects the number of rows in the AP (the AP stores 2 words per row, so the number of rows in the AP is ${sequence\ length}/{2}$). The bit-serial operation depends on precision, while the word-parallel operation on AP row count. If the row count is insufficient, bit-serial overhead offsets parallelism, reducing the AP's advantage. In addition, we characterize Softmax's runtime proportion of Llama2-7b on A100 (shown in Fig. \ref{runtime_breakdown}): For sequence lengths below $1024$ FP Softmax contributes to up to $3.34\%$ only. For sequence lengths greater than $1024$, the Softmax proportion increases to up to $38\%$, emphasizing the need to accelerate Softmax for larger sequence lengths (so the acceleration significantly impacts LLMs' overall execution time). Figs. \ref{latency} (a), (b), and (c) reveal that as sequence length increases between $1024$ and $4096$, and across all batch sizes, the AP's latency savings compared to A100(RTX3090) range respectively between $1.06\times$($1.02\times$) and $6.7\times$($12.58\times$). The $6.7\times$ Softmax speedup reduces the overall execution time of Llama2-70b by $10.71\%$ for a sequence length of $4096$ (Amdahl's law). 

\begin{figure}[t!]
\centering
{\includegraphics[width=1\columnwidth]
{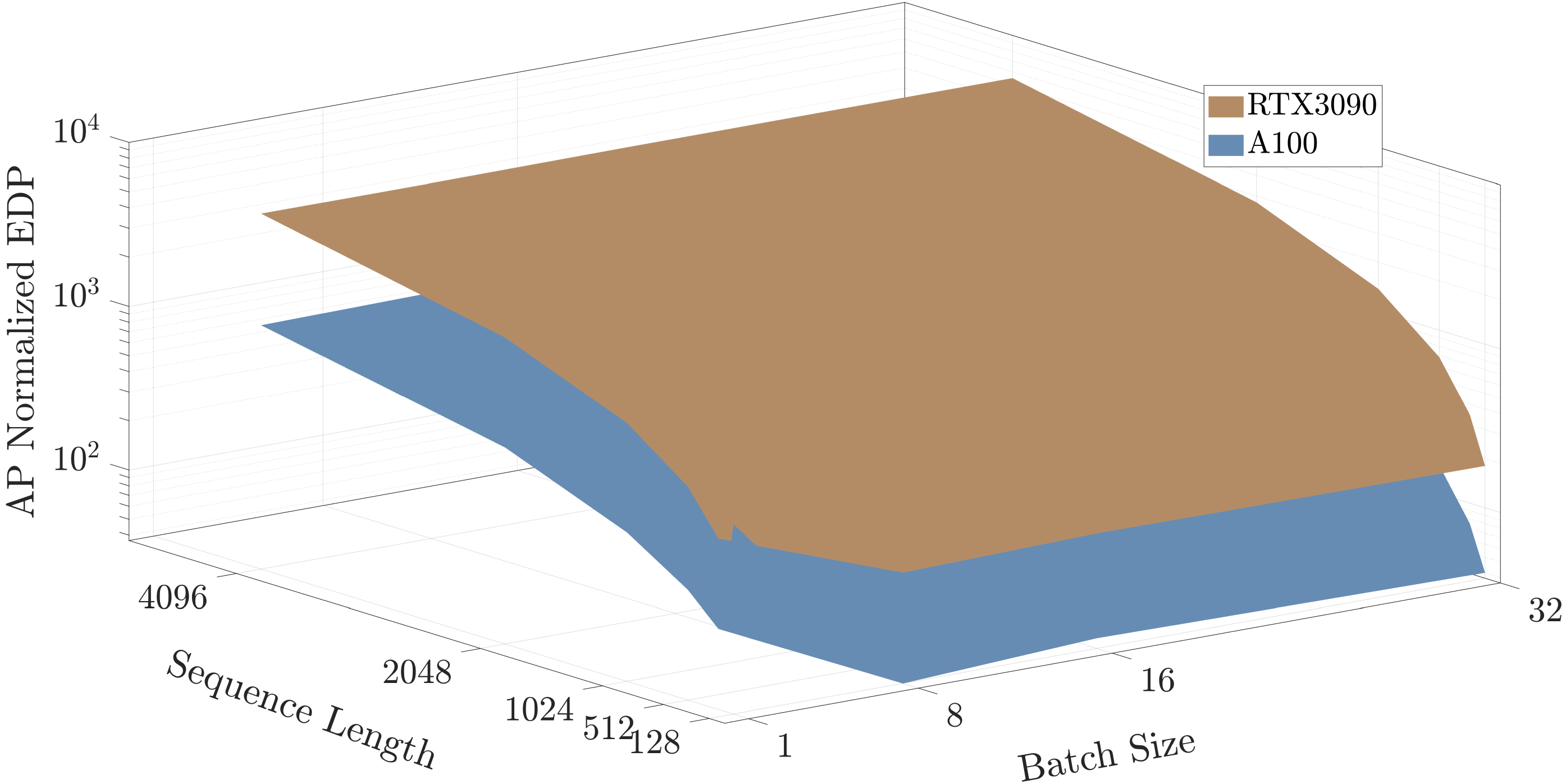}}
\caption{AP normalized energy-delay product vs sequence length and batch size for Llama2-13b.}
\label{edp}
\vspace{0.3in}
\end{figure}

To better understand the energy-latency trade-offs, we plot the normalized Energy Delay Product (EDP) in Fig. \ref{edp} for Llama2-13b. We do not include the EDP plots for Llama2-7b and Llama2-70b because the trends are similar to Llama2-13b and for space limitations. We see that the normalized EDP is always greater than 1, i.e. the AP has the lowest (the best) EDP for all sequence lengths, batch sizes, and models. This means that even when the GPUs have lower latency than the AP, the energy savings beat the latency losses, making the AP a suitable candidate for accelerating LLMs. Table \ref{bestcaseedp} reports the highest reported EDP ratio between the GPUs and APs. In the best case scenarios, the A100's EDP is $(1.07*10^3)\times$, $(1.2*10^3)\times$, and $(2.1*10^3)\times$ greater than the AP's for Llama2-7b, Llama2-13b, and Llama2-70b respectively, and the RTX3090's EDP is $(4.4*10^3)\times$, $(5.5*10^3)\times$, and $(8.8*10^3)\times$ greater than the AP's for the 3 Llama models. The ratios are $4\times$ higher with A100 compared to RTX3090, which is expected since A100 is more efficient. The highest EDPs are reported for a sequence length of 4096, and a batch size ranging between 8 and 32. The area occupied by APs to support Softmax acceleration for Llama2-7b, Llama2-13b, and Llama2-70b is $0.64mm^2$, $0.81mm^2$, $1.28mm^2$ respectively.

We note that while we present our results for Llama2 models, these results won't change for Llama3/Llama3.1 with similar Llama2 parameters (i.e. 70B) because they are structurally similar to Llama2. Additionally, our results assume standalone APs that are handling the Softmax acceleration. In a real system, the AP can be tightly coupled with CPU/GPU, whereby the CPU/GPU offloads the Softmax computation onto the AP, while it handles the remaining LLM computations. The offloading costs and exact architectural integration mechanism are outside the scope of this work.

\begin{table}[t!]
\caption{Highest EDP ratios for Llama models.}
\label{bestcaseedp}
\centering
\resizebox{\columnwidth}{!}{
\begin{tabular}{|c|c|c|c|c|c|}
\hline
\multicolumn{3}{|c|}{\textbf{Highest} $\mathbf{EDP_{A100}/EDP_{AP}}$} & \multicolumn{3}{c|}{\textbf{Highest} $\mathbf{EDP_{RTX3090}/EDP_{AP}}$} \\ \hline
\textbf{Llama2-7b} & \textbf{Llama2-13b} & \textbf{Llama2-70b} & \textbf{Llama2-7b} & \textbf{Llama2-13b} & \textbf{Llama2-70b} \\ \hline
1068 & 1191 & 2091 & 4421 & 5524 & 8851 \\ \hline
\end{tabular}
}
\vspace{-0.15in}
\end{table}

\begin{table}[b!]
\vspace{0.1in}
\caption{Comparison with related works.}
\label{comparison}
\centering
\resizebox{\columnwidth}{!}{ 
\begin{tabular}{|c|c|c|c|c|}
\hline
\textbf{Method} & \textbf{Softmax Approx.} &\textbf{Process} & \textbf{\begin{tabular}[c]{@{}c@{}}Max Freq.\\ (MHz) \end{tabular}} & \textbf{\begin{tabular}[c]{@{}c@{}}Optimum Energy\\ Per Op (pJ)\end{tabular}} \\ \hline
ConSmax~\cite{liu2024consmax} & Learnable LUTs  &16nm & 1250 & 0.2 \\ \hline
Softermax~\cite{stevens2021softermax} & \begin{tabular}[c]{@{}c@{}}Base replacement\\Online normalization\end{tabular}& 16nm & 1111 & 0.7 \\ \hline
\textbf{SoftmAP} & Integer polynomial&16nm & 1000 & $5.88 \times 10^{-3}$ \\ \hline
\end{tabular}
} 
\end{table}

\subsection{Comparison with Related Works}
We summarize the related works on accelerating Softmax under two categories: 1-) Hardware-friendly Softmax approximations and 2-) Techniques that promote Softmax parallelism.

Works in the literature in category 1-) focus on approximating the non-linear operators in Softmax, namely exponential and reciprocal, to render the deployment hardware-friendly \cite{wei2020design}. Du et al. propose an efficient hardware architecture for Softmax approximation which relies on an LUT-based exponential unit and a Maclaurin series-based natural logarithmic unit \cite{du2019efficient}. Similarly, Dong et al. propose a hardware implementation of Softmax based on a piecewise exponential LUT \cite{dong2019hardware}. However, these methods focus on Softmax in the context of DNNs and do not study the effect of the approximation on the performance of LLMs. I-BERT, the work we base our integer-only Softmax approximation on, proposes a full integer-only BERT quantization. However, it keeps the Softmax in INT32 precision (high precision) and implements the framework on the GPU (not custom hardware) using customized kernels for the non-linear operations. ConSmax \cite{liu2024consmax} is the closest to our work, as it is a software-hardware co-design methodology for a hardware-friendly alternative Softmax. ConSmax approximates Softmax using learnable parameters and proposes a corresponding custom hardware based on bitwidth-split LUT that supports mixed-precision computation. Their approach is not as easily scalable as ours as each pair of dataset/model requires retraining, which renders it not suitable in real-time scenarios. Moreover, the hardware does not inherently support mixed-precision, as the computation happens partially, and a reduction unit reads the produced partial sums and uses a chain of summations for reducing them. ConSmax is also tested on a small GPT-2 model (124M parameters) which is less complex than our Llama models (7B, 13B, and 70B parameters). Softermax \cite{stevens2021softermax} is another software-hardware methodology, whereby the custom hardware supports base replacement, low-precision Softmax computations, and calculating online normalization. Unlike our work, Softermax requires retraining, studies the impact of the approximation on BERT-Base and BERT-Large, which are smaller and less complex models compared to Llama, and relies on fixed-point formats rather than integer-only which adds hardware complexity. Table \ref{comparison} shows that SoftmAP achieves the lowest optimum energy per operation compared to ConSmax and Softermax.

Works that promote Softmax parallelism like SVD-Softmax \cite{shim2017svd}, SpAtten \cite{wang2021spatten}, and FlashAttention-2 \cite{dao2023flashattention} partition the Softmax and parallelize the execution of each partition. Our work is orthogonal to those techniques, as we can apply SoftmAP for each partition and parallelize the execution.

\vspace{-0.07in}
\section{Conclusion}
In this paper, we proposed SoftmAP, a novel software-hardware co-design methodology that implements an integer-only low-precision Softmax on the AP. SoftmAP offers up to $3$ orders of magnitude improvement in the energy-delay product compared to GPU platforms. SoftmAP presents a promising solution for making LLMs more deployable on resource-constrained devices without compromising performance. This work underscores the critical role of software-hardware co-design in advancing the efficiency of large language models.
As the demand for efficient and scalable language models continues to grow, our work underscores the importance of innovative co-design strategies in advancing the field of natural language processing.



\begin{thebibliography}{10}
\providecommand{\url}[1]{#1}
\csname url@samestyle\endcsname
\providecommand{\newblock}{\relax}
\providecommand{\bibinfo}[2]{#2}
\providecommand{\BIBentrySTDinterwordspacing}{\spaceskip=0pt\relax}
\providecommand{\BIBentryALTinterwordstretchfactor}{4}
\providecommand{\BIBentryALTinterwordspacing}{\spaceskip=\fontdimen2\font plus
\BIBentryALTinterwordstretchfactor\fontdimen3\font minus \fontdimen4\font\relax}
\providecommand{\BIBforeignlanguage}[2]{{%
\expandafter\ifx\csname l@#1\endcsname\relax
\typeout{** WARNING: IEEEtran.bst: No hyphenation pattern has been}%
\typeout{** loaded for the language `#1'. Using the pattern for}%
\typeout{** the default language instead.}%
\else
\language=\csname l@#1\endcsname
\fi
#2}}
\providecommand{\BIBdecl}{\relax}
\BIBdecl

\bibitem{zhao2023survey}
W.~X. Zhao, K.~Zhou, J.~Li, T.~Tang, X.~Wang, Y.~Hou, Y.~Min, B.~Zhang, J.~Zhang, Z.~Dong \emph{et~al.}, ``A survey of large language models,'' \emph{arXiv preprint arXiv:2303.18223}, 2023.

\bibitem{bubeck2023sparks}
S.~Bubeck, V.~Chandrasekaran, R.~Eldan, J.~Gehrke, E.~Horvitz, E.~Kamar, P.~Lee, Y.~T. Lee, Y.~Li, S.~Lundberg \emph{et~al.}, ``Sparks of artificial general intelligence: Early experiments with gpt-4,'' \emph{arXiv preprint arXiv:2303.12712}, 2023.

\bibitem{epoch2023pcdtrends}
\BIBentryALTinterwordspacing
E.~AI, ``Parameter, compute and data trends in machine learning,'' 2024, accessed: 2024-05-03. [Online]. Available: \url{https://epochai.org/data/epochdb/visualization}
\BIBentrySTDinterwordspacing

\bibitem{xu2024survey}
M.~Xu, W.~Yin, D.~Cai, R.~Yi, D.~Xu, Q.~Wang, B.~Wu, Y.~Zhao, C.~Yang, S.~Wang \emph{et~al.}, ``A survey of resource-efficient llm and multimodal foundation models,'' \emph{arXiv preprint arXiv:2401.08092}, 2024.

\bibitem{kaplan2020scaling}
J.~Kaplan, S.~McCandlish, T.~Henighan, T.~B. Brown, B.~Chess, R.~Child, S.~Gray, A.~Radford, J.~Wu, and D.~Amodei, ``Scaling laws for neural language models,'' \emph{arXiv preprint arXiv:2001.08361}, 2020.

\bibitem{kundu2024efficiently}
A.~Kundu, F.~Lim, A.~Chew, L.~Wynter, P.~Chong, and R.~D. Lee, ``Efficiently distilling llms for edge applications,'' \emph{arXiv preprint arXiv:2404.01353}, 2024.

\bibitem{li2024large}
J.~Li, J.~Xu, S.~Huang, Y.~Chen, W.~Li, J.~Liu, Y.~Lian, J.~Pan, L.~Ding, H.~Zhou \emph{et~al.}, ``Large language model inference acceleration: A comprehensive hardware perspective,'' \emph{arXiv preprint arXiv:2410.04466}, 2024.

\bibitem{wang2024model}
W.~Wang, W.~Chen, Y.~Luo, Y.~Long, Z.~Lin, L.~Zhang, B.~Lin, D.~Cai, and X.~He, ``Model compression and efficient inference for large language models: A survey,'' \emph{arXiv preprint arXiv:2402.09748}, 2024.

\bibitem{xiao2023smoothquant}
G.~Xiao, J.~Lin, M.~Seznec, H.~Wu, J.~Demouth, and S.~Han, ``Smoothquant: Accurate and efficient post-training quantization for large language models,'' in \emph{International Conference on Machine Learning}.\hskip 1em plus 0.5em minus 0.4em\relax PMLR, 2023, pp. 38\,087--38\,099.

\bibitem{dettmers2024qlora}
T.~Dettmers, A.~Pagnoni, A.~Holtzman, and L.~Zettlemoyer, ``Qlora: Efficient finetuning of quantized llms,'' \emph{Advances in Neural Information Processing Systems}, vol.~36, 2024.

\bibitem{yao2022zeroquant}
Z.~Yao, R.~Yazdani~Aminabadi, M.~Zhang, X.~Wu, C.~Li, and Y.~He, ``Zeroquant: Efficient and affordable post-training quantization for large-scale transformers,'' \emph{Advances in Neural Information Processing Systems}, vol.~35, pp. 27\,168--27\,183, 2022.

\bibitem{lin2023awq}
J.~Lin, J.~Tang, H.~Tang, S.~Yang, X.~Dang, and S.~Han, ``Awq: Activation-aware weight quantization for llm compression and acceleration,'' \emph{arXiv preprint arXiv:2306.00978}, 2023.

\bibitem{li2024fastefficient2bitllm}
\BIBentryALTinterwordspacing
J.~Li, J.~Xu, S.~Li, S.~Huang, J.~Liu, Y.~Lian, and G.~Dai, ``Fast and efficient 2-bit llm inference on gpu: 2/4/16-bit in a weight matrix with asynchronous dequantization,'' 2024. [Online]. Available: \url{https://arxiv.org/abs/2311.16442}
\BIBentrySTDinterwordspacing

\bibitem{pandey2023softmax}
N.~P. Pandey, M.~Fournarakis, C.~Patel, and M.~Nagel, ``Softmax bias correction for quantized generative models,'' in \emph{Proceedings of the IEEE/CVF International Conference on Computer Vision}, 2023, pp. 1453--1458.

\bibitem{stevens2021softermax}
J.~R. Stevens, R.~Venkatesan, S.~Dai, B.~Khailany, and A.~Raghunathan, ``Softermax: Hardware/software co-design of an efficient softmax for transformers,'' in \emph{2021 58th ACM/IEEE Design Automation Conference (DAC)}.\hskip 1em plus 0.5em minus 0.4em\relax IEEE, 2021, pp. 469--474.

\bibitem{liu2024consmax}
S.~Liu, G.~Tao, Y.~Zou, D.~Chow, Z.~Fan, K.~Lei, B.~Pan, D.~Sylvester, G.~Kielian, and M.~Saligane, ``Consmax: Hardware-friendly alternative softmax with learnable parameters,'' \emph{arXiv preprint arXiv:2402.10930}, 2024.

\bibitem{karami2024nongemm}
R.~Karami, H.~Kota, S.-C. Kao, and H.~Kwon, ``Nongemm bench: Understanding the performance horizon of the latest ml workloads with nongemm workloads,'' \emph{arXiv preprint arXiv:2404.11788}, 2024.

\bibitem{krikelis1994associative}
A.~Krikelis and C.~C. Weems, ``Associative processing and processors,'' \emph{Computer}, vol.~27, no.~11, pp. 12--17, 1994.

\bibitem{fouda2022memory}
M.~E. Fouda, H.~E. Yant{\i}r, A.~M. Eltawil, and F.~Kurdahi, ``In-memory associative processors: Tutorial, potential, and challenges,'' \emph{IEEE Transactions on Circuits and Systems II: Express Briefs}, vol.~69, no.~6, pp. 2641--2647, 2022.

\bibitem{llama2}
H.~Touvron, L.~Martin \emph{et~al.}, ``Llama 2: Open foundation and fine-tuned chat models,'' \emph{arXiv preprint arXiv:2307.09288}, 2023.

\bibitem{zhang2022opt}
S.~Zhang, S.~Roller, N.~Goyal, M.~Artetxe, M.~Chen, S.~Chen, C.~Dewan, M.~Diab, X.~Li, X.~V. Lin \emph{et~al.}, ``Opt: Open pre-trained transformer language models,'' \emph{arXiv preprint arXiv:2205.01068}, 2022.

\bibitem{brown2020language}
T.~Brown, B.~Mann, N.~Ryder, M.~Subbiah, J.~D. Kaplan, P.~Dhariwal, A.~Neelakantan, P.~Shyam, G.~Sastry, A.~Askell \emph{et~al.}, ``Language models are few-shot learners,'' \emph{Advances in neural information processing systems}, vol.~33, pp. 1877--1901, 2020.

\bibitem{chowdhery2023palm}
A.~Chowdhery, S.~Narang, J.~Devlin, M.~Bosma, G.~Mishra, A.~Roberts, P.~Barham, H.~W. Chung, C.~Sutton, S.~Gehrmann \emph{et~al.}, ``Palm: Scaling language modeling with pathways,'' \emph{Journal of Machine Learning Research}, vol.~24, no. 240, pp. 1--113, 2023.

\bibitem{reid2024gemini}
M.~Reid, N.~Savinov, D.~Teplyashin, D.~Lepikhin, T.~Lillicrap, J.-b. Alayrac, R.~Soricut, A.~Lazaridou, O.~Firat, J.~Schrittwieser \emph{et~al.}, ``Gemini 1.5: Unlocking multimodal understanding across millions of tokens of context,'' \emph{arXiv preprint arXiv:2403.05530}, 2024.

\bibitem{yantir2018efficient}
H.~E. Yantir, \emph{Efficient acceleration of computation using associative in-memory processing}.\hskip 1em plus 0.5em minus 0.4em\relax University of California, Irvine, 2018.

\bibitem{yantir2018two}
H.~E. Yant{\i}r, A.~M. Eltawil, and F.~J. Kurdahi, ``A two-dimensional associative processor,'' \emph{IEEE Transactions on Very Large Scale Integration (VLSI) Systems}, vol.~26, no.~9, pp. 1659--1670, 2018.

\bibitem{barrett1986implementing}
P.~Barrett, ``Implementing the rivest shamir and adleman public key encryption algorithm on a standard digital signal processor,'' in \emph{Conference on the Theory and Application of Cryptographic Techniques}.\hskip 1em plus 0.5em minus 0.4em\relax Springer, 1986, pp. 311--323.

\bibitem{kim2021bert}
S.~Kim, A.~Gholami, Z.~Yao, M.~W. Mahoney, and K.~Keutzer, ``I-bert: Integer-only bert quantization,'' in \emph{International conference on machine learning}.\hskip 1em plus 0.5em minus 0.4em\relax PMLR, 2021, pp. 5506--5518.

\bibitem{milakov2018online}
M.~Milakov and N.~Gimelshein, ``Online normalizer calculation for softmax,'' \emph{arXiv preprint arXiv:1805.02867}, 2018.

\bibitem{wikitext}
S.~Merity, C.~Xiong \emph{et~al.}, ``Pointer sentinel mixture models,'' \emph{arXiv preprint arXiv:1609.07843}, 2016.

\bibitem{pytorch}
PyTorch, \url{https://pytorch.org}.

\bibitem{transformer}
``Transformers,'' \url{https://huggingface.co/docs/transformers/main/index}.

\bibitem{HuggingFace}
HuggingFace, 2024, \url{https://huggingface.co/}.

\bibitem{rakka2024bf}
M.~Rakka, R.~Karami, A.~M. Eltawil, M.~E. Fouda, and F.~Kurdahi, ``Bf-imna: A bit fluid in-memory neural architecture for neural network acceleration,'' \emph{arXiv preprint arXiv:2411.01417}, 2024.

\bibitem{wei2020design}
Z.~Wei, A.~Arora, P.~Patel, and L.~John, ``Design space exploration for softmax implementations,'' in \emph{2020 IEEE 31st International Conference on Application-specific Systems, Architectures and Processors (ASAP)}.\hskip 1em plus 0.5em minus 0.4em\relax IEEE, 2020, pp. 45--52.

\bibitem{du2019efficient}
G.~Du, C.~Tian, Z.~Li, D.~Zhang, Y.~Yin, and Y.~Ouyang, ``Efficient softmax hardware architecture for deep neural networks,'' in \emph{Proceedings of the 2019 on Great Lakes Symposium on VLSI}, 2019, pp. 75--80.

\bibitem{dong2019hardware}
X.~Dong, X.~Zhu, and D.~Ma, ``Hardware implementation of softmax function based on piecewise lut,'' in \emph{2019 IEEE International Workshop on Future Computing (IWOFC}.\hskip 1em plus 0.5em minus 0.4em\relax IEEE, 2019, pp. 1--3.

\bibitem{shim2017svd}
K.~Shim, M.~Lee, I.~Choi, Y.~Boo, and W.~Sung, ``Svd-softmax: Fast softmax approximation on large vocabulary neural networks,'' \emph{Advances in neural information processing systems}, vol.~30, 2017.

\bibitem{wang2021spatten}
H.~Wang, Z.~Zhang, and S.~Han, ``Spatten: Efficient sparse attention architecture with cascade token and head pruning,'' in \emph{2021 IEEE International Symposium on High-Performance Computer Architecture (HPCA)}.\hskip 1em plus 0.5em minus 0.4em\relax IEEE, 2021, pp. 97--110.

\bibitem{dao2023flashattention}
T.~Dao, ``Flashattention-2: Faster attention with better parallelism and work partitioning,'' \emph{arXiv preprint arXiv:2307.08691}, 2023.

\end{thebibliography}
\end{document}